\documentclass{arXiv.sig-alternate}

\usepackage{color}

\begin{document}


\title{Reference and Structure of Software Engineering Theories}
%
\numberofauthors{1} 
\author{
\alignauthor
Andr\'{e}s Silva\\
       \affaddr{GIB research group}\\
       \affaddr{Universidad Polit\'{e}cnica de Madrid}\\
       \affaddr{28660 Boadilla del Monte, Spain}\\
       \email{asilva@fi.upm.es}
}

\maketitle
\begin{abstract} 
This paper tries to contribute towards the solution of an important question raised in \cite{Hannay}: What is a Software Engineering (SE) specific theory? Which are the main features of a theory that is endemic to SE? In this paper we will use ``theory'' as the term is used in traditional sciences. Other uses of the term ``theory'' are discussed. Finally, we propose to focus on the {\it reference class} and on the {\it structuring} of SE theories as a basis for further progress. 
\end{abstract}


\section{Introduction} 

Semantics is the field concerned mainly with meaning and true \cite{Bunge}. In particular, the semantics of science are concerned with the sense, the reference and the truth of those scientific sentences which, expressed in different formalisms, constitute scientific theories. In factual (i.e. non-formal) sciences, theories are well-structured bodies of sentences about real objects, their properties and their processes. In formal sciences, on the other hand, the sentences in a theory refer to conceptual items, like the triangle, or $\pi$. The discussion in this paper will be related to the reference of factual scientific theories, leaving the sense and truth value out of scope. 

For example, the Kinetic Theory of Gases (KTG) is a body of sentences, expressed in a mathematical formalism, that refer to molecules of a gas, describing their movements and other features from which high-level properties are derived (gas temperature, pressure, volume, etc.). It is obvious that KTG does not deal with electromagnetic fields or, in other words, electromagnetic fields do not constitute the reference of KTG.  This seems pretty obvious in this example, but is not so obvious in other fields of science and, in particular, in novel fields of study, as discussed later.

The objects that are the referents of a particular sentence constitute its {\em reference class}. However, we must take into account that, in science, reference classes may be hypothetical or non-existent. For instance, the luminous ether was the subject of many a theory in the 19th Century. It would not be correct to say that theories about the luminous ether had no reference. Similarly, it would be incorrect to say that theories about dinosaur disappearance have no reference class. It is more precise to say that the {\em extension} of those theories happens to be empty, but their intended (i.e. hypothetical) reference class is not. In fact, if the extension is not yet determined, at least the theory can guide researchers in the search for evidences in support of its existence, like fossils. This process also takes place in Theoretical Physics, when a particle is first postulated. 

In brief, any sentence that is part of a scientific theory $(i)$ has a reference class; $(ii)$ can be supported (or not) by a set of empirical findings and $(iii)$ a truth value can be attributed to it. However, it is better not to confuse those things. The reference class of a theory is not the same as its empirical proofs. These are, in some cases, a small subset of the reference class or, if not, they can be connected to the reference class through a chain of auxiliary theories, like those about the behavior of observation instruments \cite{Bunge}. What is important is that, for a scientific theory, having a nonempty reference class is a necessary (but not sufficient) condition for being empirically testable. In consequence, before testing a theory we first should know what the theory is talking about, and for choosing between competing theories we must be sure that they are about the same things. 

With this in mind, looking to the field of Software Engineering (SE) and looking at some of its most widely used theoretical sentences \cite{Cain}, we could ask: are Conway's law and Brooks' law talking about the same things? In other words, are they competing? Are they mutually supportive or are they completely independent? Empirical proofs of one of those laws could be used as empirical proofs of the other? Is Parnas's principle of information hiding a competing theory of Conway's law? If they are not competing, do they overlap? Could both of them be part of a same umbrella theory? Also, it has been pointed out that Dijkstra's famous attack on the GOTO statement is a theory of cognitive limits \cite{Pontus}. Then, should human minds belong to the reference class of that theory? If this is true, in the search for empirical evidence perhaps we  should pay attention to what psychologists have to say, in addition to computer scientists. 

In consequence, the semantics of science are very relevant for, and should precede, the methodology of science \cite{Bunge}. The position advocated in this paper is that it would be very helpful, before developing SE-specific theories, to first clarify their hypothetical reference class. This is not an empty exercise, as it would help to define the universe of discourse and the subject matter of those theories. Additionally, from a methodological viewpoint, this is a helpful exercise for identifying competing theories, for searching empirical evidence that supports them and for finding to which extent some theories give account of particular facts (for instance, empirical evidence that was not supported by Newton's gravity was supported by Einstein's). But first let's briefly discuss what we mean when we say that we miss theories for SE.

\section{What do we miss in SE?}

A desire for theory has been expressed many times in the literature, but the use of particular theories as a background to many SE published research is negligible \cite{Hannay}. However, there are two issues related to the use of ``theory'' that we would like to clarify\footnote{A third issue happens in lay language, when ``theory'' is misused as a synonym of ``hypothesis''.}.

\subsection{Do we miss ``theoretical content'' in SE?}

It is important to point out the distinction between theory and ``theoretical content''. This is a distinction widely used by students and teachers. They regard as ``theoretical'' a class where the teacher talks about the syntax of a programming language, for instance. On the contrary, a class where students develop actual programs would be considered ``practical''. This usage of ``theory'' is mainly to differentiate practical studies from those which are not. In this regard, ``Film Theory'' embraces the studies about films and filmmaking, but is not related to the actual task of film production. ``Feminist Theory'' analyzes gender inequality. The ``Theory of Poetry'' embraces studies about poetry as a literary form. There is even an ``Architectural Theory'', about the artistical and social dimension of architecture, with no direct relation to the underlying theories of structures or materials. Even in formal sciences ``theory'' sometimes points to collections of related ``theoretical content'', like in ``game theory'' or ``network theory'', which designate broad areas.


This usage of ``theory'' does not happen in well established factual sciences, as these sciences strongly emphasize the need for empirical data in support (or against) any theory. Clearly ``Einstein's theory'' is not just a set of studies on the speed of light and the space-time curvature: it has strong empirical support. Scientific theories can even be classified according to the degree of empirical support they have, which would be difficult to do with film or feminist ``theories''. Additionally, from a scientific theory it is always possible to derive statements that can empirically tested. Again, that would be very difficult to do for ``film theory'' and even for ``game theory'' alone, without the addition of further correspondences between the theory and the real world. 

It is not difficult, in SE, to find ``theory'' in this sense of ``theoretical content''. The existence of bodies of knowledge and classical textbooks on the subject of SE could be used as a proof. However, theories in the scientific sense, who explain different aspects of the SE world and are supported by a strong set of empirical evidence, are still missed.

\subsection{Do we miss ``laws'' in SE?}

In \cite{Pontus} it is said that, despite some opinions, ``theory is abundant'' in SE, and examples of scientific theory are provided (the logarithmic curve of change costs, information hiding, etc.) but, as the authors show, they are very small and casual. We claim that those examples are actually more akin to scientific {\em laws} than to scientific {\em theories}. This situation is also reflected in \cite{Hannay} where it is shown that theory use is very local in empirical SE research, with a lack of theory building efforts. Again, we claim that the reason is because that research is more law- than theory-oriented. 

Scientific theories go beyond scientific laws, as these are narrower in scope, usually circumscribed to a particular cause-consequence relationship. For example, Boyle's law explains the relationship between gas pressure and volume. Laws can be part of a theory but they are not a theory by themselves. In fact, it is the great context of a theory what explains a scientific law. In this sense, Boyle's {\em Law} can be derived from the Kinetic {\em Theory} of Gases. Another example is Planck's {\em Law}, about black body radiation, that is included in, and explained by, Quantum {\em Theory}.

The fact that in SE there is a set of well-established laws must be welcome, but the issue about the lack of SE theories still remains. A general theory of SE, even a wrong theory, needs to look farther than a mere set of laws, however precise and useful they are. If we pay attention to the most successful scientific theories, we will find very good structure and internal organization, where laws are pieces subordinated to the overall architecture of the theory. We will return later to this issue of theory structuring. In resume:
\begin{itemize}
\item Yes, there is theoretical content in SE, like there is in the fields of literature, film, etc. Is that enough? 
\item Yes, there are laws in SE that explain cause-consequence relationships, well-known by the SE community of researchers and practitioners. Are those laws enough?
\end{itemize}
So the question of "where's the theory for SE?" \cite{Pontus} still remains, but we hope to have scoped that question a little.

\section{The referents of a SE theory}

To illustrate the problem of finding the reference class of a sentence, let's take ``all swans are white'' or $\forall (x)$ $Swan(x)$ $\Rightarrow$ $White(x)$. Leaving aside the truth or falsity of the sentence, it looks like a sentence about swans. However, it is equivalent to $\forall (x)$ $\neg White(x)$ $\Rightarrow$ $\neg Swan(x)$ so finding any non-white object (an apple, for instance) that is not a swan would seem to provide empirical support. It is dubious that such piece of empirical evidence would be accepted by an ornithologist. Actually, in an ornithology context, the sentence would be regarded as a sentence about birds and nothing else. Usually, in scientific contexts, the referents of a theory are better defined than in lay contexts, but this is not always the case and is not a trivial matter.

In Physics, Quantum Theory is about microparticles and microsystems, and that is how scientifics use it in their daily work at CERN or other research facilities. However, sometimes they affirm that the theory is about microsystems plus measurement apparatus plus conscious observers. According to the prominent Physicist E. Wigner, historically, when physics tried to encompass the microscopic world ``it was not possible to formulate the laws of quantum mechanics in a fully consistent way without reference to the consciousness'' \cite{Wigner}. Despite that, researchers in the physics field proceed everyday without paying attention to consciousness in the formulas they use, probably because there is nothing on those formulas that points to consciousness at all (in other words, consciousness is not in the reference class of the formulas). Heisenberg said that Quantum Theory is not about Nature, but about the knowledge of Nature instead \cite{Bunge}. However, the equations in the theory do not contain variables that point to observers or knowledgeable beings or apparatus and, in fact, those theories are true even for particles that are in far-away galaxies. Other controversies arise in biology: is Evolution Theory about individuals, species, populations, genes or all of them at the same time? Is Relativity Theory also about observers? What about Control Theory? Control Theory can be applied to machines and animals so, what's in its reference class? If setting the referents of a theory is controversial in well-established fields, SE is not an exception. Formally expressed, the basic question of finding what a theory for SE should talk about is:
\begin{quotation}
Which real-world objects, properties and processes should be in the reference class of a Software Engineering Theory?
\end{quotation}
This is a question that is independent of the true or falsity of potential theories. A false theory about electrical charges still has electrically charged bodies in its reference class. In fact, having false theories and proving (empirically) that they are false, is always welcomed. Competing theories must exist in order to choose the better ones, but only theories that share elements in their reference classes are competitive. A theory about protein synthesis does not compete with a theory about unemployment, unless someone discovers a third theory that connects protein synthesis with unemployment. In SE, for instance, does a theory about software patterns relate to theories about market share? Is this even a legitimate question or should be out of the scope of any potential SE theory?

As a first step, it looks easy to point out to real objects that apparently should not be part of the reference class. On the one hand, nobody would seriously think that motorcycles, minerals or proteins should be part of it. On the other hand, source code, requirements documents or test cases that should possibly be there. But what about contracts, CPUs or salaries? What about organization structures or project schedules? A prominent feature of the potential elements to be included in the reference class for a SE theory is that they are created by humans, and do not exist in nature, which takes us to the following section.

\section{A science of the artificial}

Wrongly or not, we are taking scientific theories as our ``role model''. However, both the NAS and the AAAS define scientific theories by remarking that they make reference to the natural world. So, how can SE be the subject of a scientific theory when it deals with human produced artifacts? For finding an answer we should pay attention to the studies of H. A. Simon on the Sciences of the Artificial \cite{Simon}. These sciences deal with the practical issues of conceiving and constructing new and useful artifacts, not with the study of pre-existing objects like sciences of nature do. Hence, according to our discussion on the reference class of scientific theories, a theory about engineering should contain statements not only about the software itself but also about the humans and the activities they are engaged in during the construction process.

Let's  consider first what happens in other engineering fields, by taking Electrical Engineering (EE) as the paradigm. EE rests, among others, on Maxwell's Theory, which is about electrically charged bodies and fields. However, a theory about electrical charges is not an EE theory, because EE is what electrical engineers {\em do}, which is mainly to design, construct, fabricate and test electrical equipment, given a set of limited resources, tools, knowledge, etc.  Maxwell's equations do not care about the process of building and testing new devices (i.e. those elements are not in its reference class). Which features, then, should have a SE theory that is not a merely a theory of software? In order to answer this, we propose this distinction:
\begin{itemize}
\item Laws and Theories whose reference class includes the objects that belong to the subject matter of any engineering (bodies, fields, materials, engines, etc.) will be called Foundation Laws or Theories (briefly, F-Laws or F-Theories). Example: Boyle's Law, Hooke's Law, Maxwell's Theory, KTG, etc. 
\item Laws and Theories whose reference class includes the engineering practice (from the subject matter to the people involved, the designs and its evolution, the testing processes, etc.) will be called Engineering Laws or Theories (E-Laws or E-Theories). Examples: Manufacture Scheduling Theory, Brooks's Law, etc. 
\end{itemize}

Simplifying, F-Theories belong to the Sciences of the Natural and E-Theories to the Sciences of the Artificial. Traditional engineering fields are based upon a solid basis of F-Theories but they are not specially rich on E-theories or, at least, on particular E-theories specially fitted to each engineering subdomain. These engineering practices rely on a well-developed set of domain-independent E-theories (like scheduling theory or planning theory) which can be adapted to the particularities of, for instance, civil or naval engineering. For instance, in EE, despite the wide corpus of well-tested F-theories in use, it is not easy to find an E-Theory that is EE-specific, perhaps because the need of such a theory was, historically, less urgent in EE that it is in SE. Other engineering domains were traditionally more prone to task decomposition and parallelization; not so in SE, as the Software Crisis clearly showed. This is why the debunking of the person-month myth \cite{Brooks} never happened until the second half of the 20th Century, in a software context, a fact whose importance is usually underrated. It is the opinion of this author that SE should not envy other engineerings, because they are also not specially rich in E-Theories. However, the need for an E-Theory is more urgent in SE due to its characteristics. This is what gives SE a special place in the landscape of engineering.

In the case of software we may regard the corpus of Computer Science theories (programming language theory, complexity theory, algorithmic, etc.) as our F-Theories. We have also some E-Laws but, and this is at the core of the {\em position} of this paper, finding E-Theories is more urgent in SE than in any other fields, due to special characteristics of software development. In the same way, other engineering fields were traditionally less interested in E-Theories than we are (traditionally, of course, because they are also incorporating software in their domains, and start suffering the same problems that plagued SE). SE, from its early days, is known by the highly interaction that is present among subtasks and by the sheer importance of human factors and management issues. I suggest the reader to compare SE textbooks with those in other engineering fields, and observe the amount of material devoted to human factor and organizational issues in SE books. In fact, it could be said that it was this human-factor awareness what actually gave birth to the SE discipline, as a means to break free from the tar pit \cite{Brooks}. As  this human and organizational factor is so relevant, we should incorporate it in our potential theories.


In other words, F-Theories and E-laws are welcome, but what we need in SE are E-Theories. This paper suggest to start with the popular {\em three P's of SE}, or PPP, as a first approximation towards establishing the reference class of an SE E-Theory. These three P's are:
\begin{itemize}
\item People: Engineers, users, testers and, possibly, other stakeholders involved in the SE effort.
\item Product: Artifacts generated during a SE effort, including the computer programs (i.e., not only the algorithms, understood as a mathematical construct).
\item Process: The different steps carried out by the People in order to transform the Product through different stages which, following \cite{Simon}, should include the search for design alternatives, evaluation of alternatives, representations and their transformation, among others.
\end{itemize}
In this proposal, an E-Theory of SE must refer to the three Ps {\em together}. A theory about one, or two, of the P's would be a partial SE E-theory (but could be part of a wider theory). We believe  this PPP proposal is, at least, a basis to trigger a discussion on the subject matter of SE theory.

\section{Theories have structure}

Scientific theories are constituted by well organized sentences. They are not just a random set of consistent affirmations about some real world domain. An example of a theory with a deep and well-developed structure is the Theory of Evolution in its Modern Synthesis, where Darwinism explains phenomena and Genetics explains the reasons behind those phenomena. Another example is the KTG: macro and micro level are in perfect harmony and what happens at the molecular level explains what happens at the macro level. In the software world it would be wise to ask if, in addition of searching for new E-theories, perhaps it would be wise to better structure the knowledge we already have, mostly under the form of F-Theories and E-laws. In this way, and in analogy to the KTG, we could relate the macro and micro landscapes. Examples of research questions pointing to this goal may be: does object orientation increase the performance of the usability evaluation process? If yes, or not, which are the reasons? Which software architectures do improve user involvement in the development process, if any? Which is the relationship, if there is one, between the structure of development teams and non-functional requirements like robustness or availability? If there is no relationship, why? Other examples are given in \cite{Hannay}:  ``expert familiarity of design patterns when designing safety-critical systems in a short time-to-market strategy, leads to better reliability'', or ``comment lines describing design patter usage will increase programmer performance''. 

These are examples of questions that could be explained by good E-theories in SE, or that could give birth to such E-theories, if more effort is put into their study and validation. And, if these questions make no sense, that could also be explained by good theory. Searching in the literature, it is possible to find works that already point in this direction, of finding explanatory connections between software-related issues that happen at different levels. In \cite{Cain} empirical evidence is provided for sentences like ``the architecture of the software alters the number of programmers that can effectively work together'', or ``If software is improperly coupled then people are improperly coupled", where ``improper'' is defined by the presence of some visible attributes like rigidity, fragility, immobility and viscosity \cite{Cain}. From their findings, predictions can be made, as ``the lack of coupling management tasks within a project will slow down development''. Independently of which was the goal of the researchers, these are good examples of sentences that link the three P's, with empirical support, being then good candidates to become the seed of, or be part of, a potential E-theory. 

\section{Conclusions}

The position of this paper is: $(i)$ Semantically: To use PPP as the reference class of prospective SE E-theories and $(ii)$ Methodologically: To structure and organize already known material. Of course, both $(i)$ and $(ii)$ go together, because the structuring of the material could be guided by its coverage of PPP. Optimistically, we could say that we already have F-laws, F-theories and E-laws, but we are in a strong need of E-theories, understood as well organized and empirically testable sets of sentences whose reference class are the People, the Process and the Product.

%

\bibliographystyle{abbrv}

\begin{thebibliography}{1}

\bibitem{Brooks}
F. P. Brooks.
\newblock{\em The Mythical Man-Month.} 
\newblock Addison-Wesley, 1975.

\bibitem{Bunge} 
M. Bunge.
\newblock{\em Treatise on Basic Philosophy: Semantics I: Sense and Reference}.
\newblock Springer, Nov. 1974.

\bibitem{Cain}
J. W. Cain and R. J. McCrindle.
\newblock An investigation into the effects of code coupling on team dynamics and productivity.
\newblock In {\em Proceedings of the 26th Annual International Computer Software and Applications Conference (COMPSAC 2002)}, pages 907--913. IEEE Computer Society, Aug. 2002 .

\bibitem{Hannay}
J. E. Hannay, D. I. K. Sjoberg, and T. Dyba.
\newblock A systematic review of theory use in software engineering experiments. 
\newblock {\em IEEE Transactions on Software Engineering}, 33(2):87--107, 2007.

\bibitem{Pontus}
P. Johnson, M. Ekstedt and I. Jacobson.
\newblock Where's the theory for Software Engineering?
\newblock {\em IEEE Software},  29(5):94--96, Sep. 2012.

\bibitem{Simon}
H. A. Simon.
\newblock{\em The Sciences of the Artificial}. 3rd ed.
\newblock MIT Press, 1996.

\bibitem{Wigner}
E. Wigner.
\newblock Remarks on the Mind-Body Question.
\newblock In {\em Quantum Theory and Measurement}. J. A. Wheeler and W. H. Zurek, eds., 
\newblock Princeton University Press, 1983.




%
%
%
%
%

\end{thebibliography}

\end{document}